\documentclass[12pt,preprint]{aastex}

\slugcomment{Submitted to ApJ Suppl.}

\shorttitle{K-line Spectra in CP Stars}
\shortauthors{Cowley, Hubrig, and Kamp}

\begin{document}


\title{An Atlas of K-line Spectra for Cool Magnetic CP Stars \\
   The Wing-Nib Anomaly (WNA)}


\author{C. R. Cowley,}
\affil{Department of Astronomy, University of Michigan,
    Ann Arbor, MI 48109-1042}
\email{cowley@umich.edu}

\author{S. Hubrig,}
\affil{European Southern Observatory, Casilla 19001, Santiago 19,
Chile}
\email{shubrig@eso.org}

\and

\author{I. Kamp}
\affil{Space Telescope Science Division of ESA, STScI, Baltimore, MD 21218}
\email{kamp@stsci.edu}



\begin{abstract}
   We present a short atlas illustrating the unusual
   Ca {\sc ii} K-line profiles in upper main sequence stars with
   anomalous abundances.
   Slopes of the profiles for 10 cool,
   magnetic chemically peculiar (CP) stars
   change abruptly at the very core, forming a deep ``nib."
   The nibs show the same or nearly the same radial velocity
   as the other atomic lines.  The near wings are generally
   more shallow than in normal stars.  In three magnetic
   CP stars, the K-lines are too weak to show this shape,
   though the nibs themselves are arguably present.  The
   Ca {\sc ii} H-lines also show deep nibs, but the profiles are
   complicated by the nearby, strong H$\epsilon$ absorption.
   The K-line structure is nearly unchanged with phase in
   $\beta$ CrB and $\alpha$ Cir.   Calculations,
   including NLTE, show that other possibilities in addition
   to chemical stratification may yield nib-like cores.

\end{abstract}



\section{Introduction}

Abundance studies of chemically peculiar (CP) stars of the
upper main sequence are usually based on classical, 
plane-parallel atmospheres.  Until recently, there was 
little evidence that this assumption would lead to 
significant errors, apart from the obvious case of spectrum 
variables.  Even in these cases, it was common to assume 
plane-parallel structure for localized regions (abundance 
patches) of the photosphere.  

Recent work on the ionization equilibrium
\cite{ryab04}
and the Balmer profiles \citep{cowl01}
of cooler CP stars in the magnetic sequence has drawn
attention to significant departures from classical 
atmospheric structure. 

Decades ago, spectroscopists had noted another indication 
of departures from a classical atmosphere in the Ca {\sc ii}
K-lines.  
\cite{baba58} describes very complicated
K-line profiles in a number of CP stars.  In a few cases,
variable profiles suggested to him
``ejection of ionized clouds or streams such as the sun
produces."   Analytical spectroscopists did little
more than note these remarks at the time.  Most of the
interest in CP star focused on their peculiar abundances
and attempts, for example, by \citet{conti65} and
\citet{vveer66} 
to explain the abundance anomalies in Am stars
by unorthodox models had been unsuccessful.

New, high-resolution observations make it appropriate to
take a closer look at the K-lines of CP stars.  We confine
ourselves to  spectra of cooler magnetic stars with minimal
rotational broadening.  We find that the K-lines of these
stars generally exhibit
a ``wing-nib anomaly'' (WNA).

\begin{figure}[t]
   \centering  
       \includegraphics[width=0.3\textwidth,angle=0,scale=1.4]{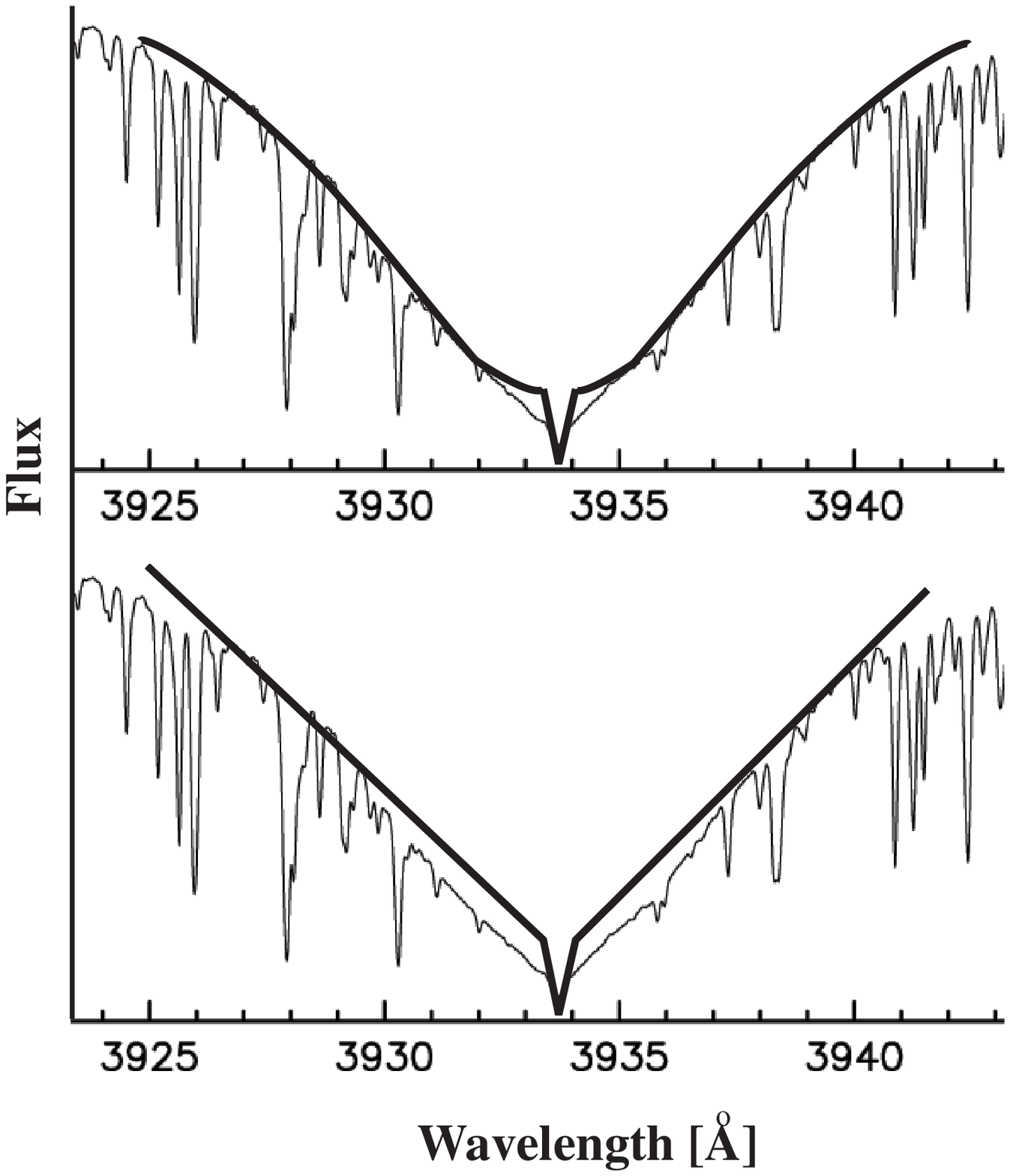}
      \caption{Schematic profiles for the cool, magnetic CP
         stars (heavy line).  Specific examples of both kinds of
         profiles are illustrated in subsequent figures.
         Vertical lines on either side of the v-shaped ``nib''
         indicate the nib width.  Numerical values are given
         in Table~\ref{tab:nibwidths} below.
         The underlying stellar
         spectrum is of the F2 V star $\sigma$ Boo.
         } 
         \label{fig:one}  
   \end{figure}

\placefigure{fig:one}

Two forms for the  WNA are shown schematically in
Fig.~\ref{fig:one}. In the upper part of the figure, the
wings show a gently changing slope, which appears to
nearly level off prior the deep minimum.  In the
lower part of the figure, the wings show an approximately
constant slope beyond the deep nib.  In both cases, the
region outside of the nib has a higher intensity than
is observed for normal stars, or the Am star HR 1353.
Individual examples are cited in the captions of figures
to follow.

The Ca II H-line also shows a sharp, deep nib, but the
structure is complicated by the presence of the strong
Balmer H$\epsilon$ line.   We shall not discuss the
H-line in this paper.

\section{Previous K-line Work}

Complications in the core of the Ca {\sc ii} K-line
(as well as the
H-line) are well known in stellar astronomy.  The most 
common phenomena are single and double reversals (e.g. 
K$_2$, and K$_3$).  These reversals are common among F-type 
stars, and may be seen, weakly, in the central cores of
several non-CP stars discussed later in the paper.
Additionally, there are numerous examples of sharp, 
interstellar components, which typically have
different radial velocities than the stellar lines.

It is likely that
\cite{babb58}  
saw the WNA among
a variety of other phenomena that influence the K-line
structure in CP stars.  However, his papers give
verbal descriptions, not detailed profiles.

An important observational step was
taken by Babel (1994).  He measured two K-line 
``equivalent widths,'' a $W_{0.3}$ within $\pm 0.3$~\AA\ of
the line center, and a $W_{10}$ within $\pm 10$~\AA.  
He showed that for a given $W_{10}$, the $W_{0.3}$ was 
smaller for Ap stars than for normal stars.  In other 
words, the core absorption of Ap stars was smaller than
for normal stars with comparable overall K-line 
strengths.  Note this does not of itself indicate the
presence of a nib.  Clearly nibs, when present, increase
the $W_{0.3}$ absorption rather than reduce it.  The
$W_{0.3}$/$W_{10}$ trend found by Babel persists in
spite of the nibs rather than because of them.

The current atlas elaborates these findings for 
the small class of Ap stars.
Our UVES spectra have twice or more typically 
four times the resolution employed by Babel.  Finally,
because we deal with profiles rather than equivalent 
widths, we show in detail how the K-line absorption in 
peculiar {\it and normal} stars differs.  

Ca K-line nibs may indeed be seen for three stars in the
papers by Babel (1992, 1994) and 
\citet{ryab02} 53 Cam, $\gamma$ Equ, and $\beta$ CrB.
None of these papers contrast the nib-structure
with the Ca K {\it profiles} of normal stars.

\section{Observations}

All spectra were obtained with the
VLT UV-Visual Echelle Spectrograph 
UVES at UT2. The major part of the spectra of magnetic CP stars
was obtained in the course of our study of chemical abundances
of CP stars (ESO programmes Nos.\ 70.D-0470,
072.D-0414, and 074.D-0392).
A few other spectra of CP stars were retrieved from
the ESO UVES archive (ESO programme No.\ 68.D-0254).
Spectra of some other stars were downloaded from the UVESPOP web
site \citep{bag04}.  
Details are summarized in
Table~\ref{tab:dates}; spectral types are
from \citet{ren91} and SIMBAD.  
The stars are all bright, and signal-to-noise
ratios greater than 200 were typical.  In a few cases
where multiple scans were available, the spectra were
averaged, weighted by the number of counts.

\placetable{tab:dates}

All spectra were observed with Dichroic standard settings at the resolving power   
of $\lambda{}/\Delta{}\lambda{} \approx 0.8\times10^5$ and 
have been reduced by 
the UVES pipeline Data Reduction Software,  
which is an evolved version of the ECHELLE context of MIDAS.  

The 13 CP stars presented here were selected on the basis
of (1) their chemical peculiarity, (2) their low values of
the projected rotational velocity
($v\cdot\sin(i)$), and (3) the availability of high-resolution,
high signal-to-noise spectra.  All stars
have effective
temperatures between approximately 6600 and 9600K.  Of 
these, 10 show the WNA. The three stars without a WNA have 
Ca K profiles that are too weak to have extended wings.  In 
the case of two of these stars, HR 5623  and HR 7552,
the stars are too hot to have strong
K-lines.  The K-line is also too weak to show the WNA in 
Przybylski's star.  Przybylski's star is the
coolest in our sample, but the K-line is weak because of 
the low calcium abundance.

\begin{figure*}
   \centering  
       \includegraphics[scale=0.8]{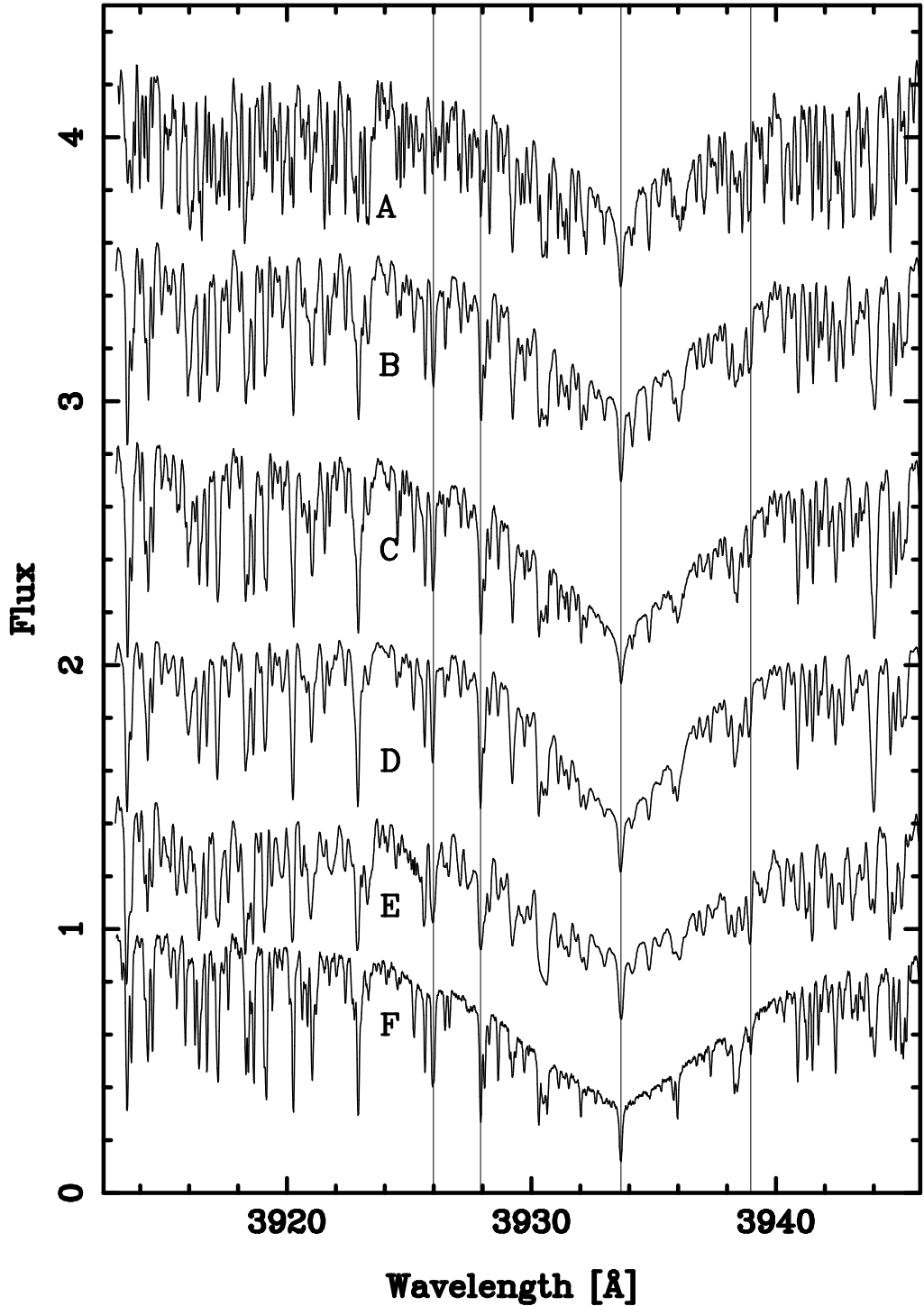}
      \caption{K-line region for A (HD 217522), B (HD 134214),
      C (HD 122970), D (HR 1217), E (33 Lib),
      \& F (10 Aql).  Spectra are normalized to a
      high point near 3915~\AA.  All but the lowest are displaced
      upward for display purposes.
      Vertical lines are at
      $\lambda\lambda$3925.99 (Fe{\sc i}-364,562),
      3927.92 (Fe {\sc i}-4),
      3933.66 (Ca {\sc ii}-1), 3938.97 (Fe {\sc ii}-190).
      Multiplet numbers \citet{moore45}
      for dominant species are indicated along with the atomic
      spectra.  All features are blended.
              }  
         \label{fig:cold}
   \end{figure*}

\placefigure{fig:cold}

\begin{figure*}
   \centering  
           \includegraphics[scale=0.9]{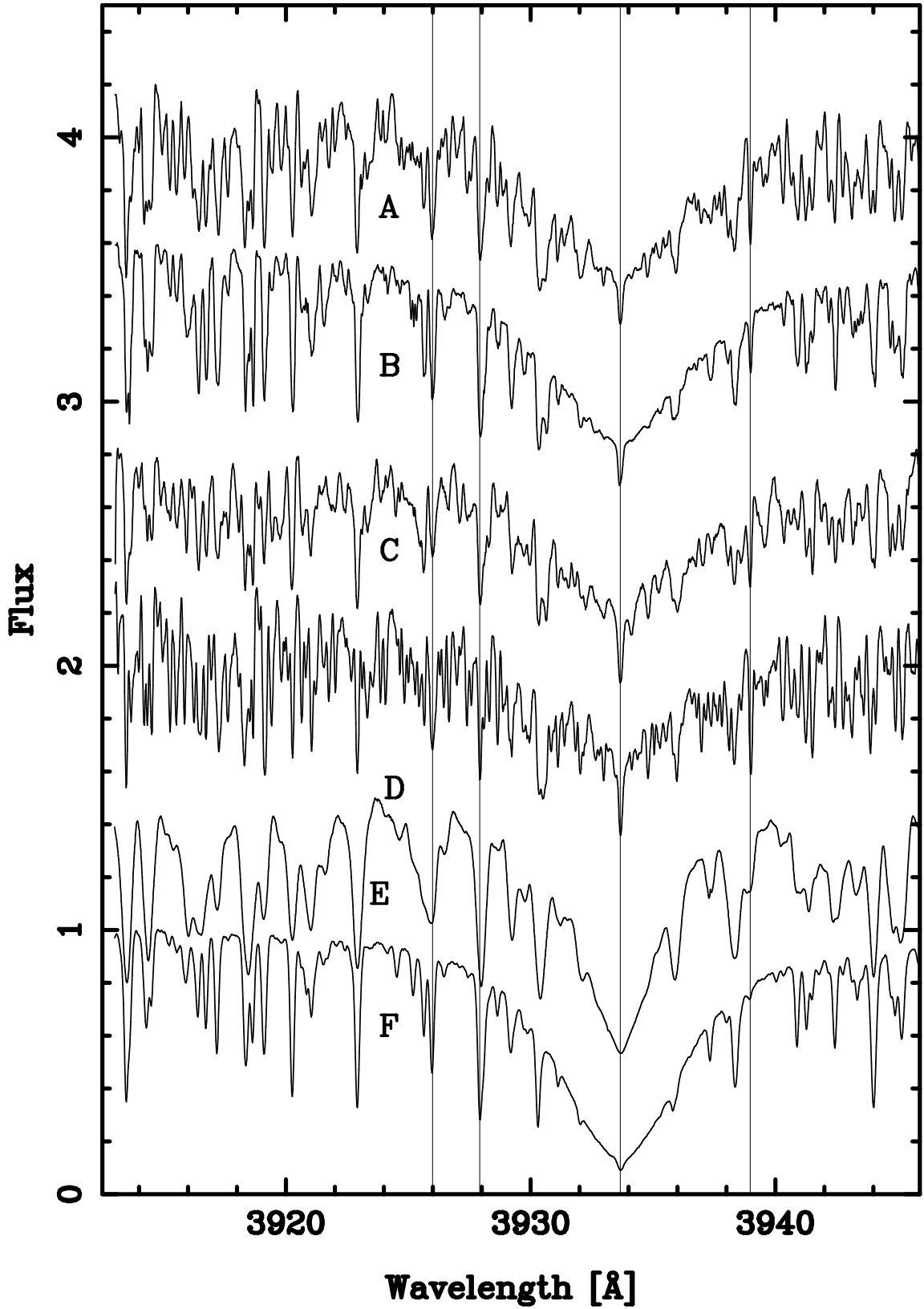}
      \caption{K-line region for A ($\beta$ CrB),
      B (HD 116114),
      C (HD 216018), D (HR 7575),
      E (HR 1353, Am),
      \& F ($\sigma$ Boo).
      Vertical lines as in previous
      figure.
              }  
         \label{fig:hot}
   \end{figure*}

\placefigure{fig:hot}

%

%

%

%

The magnetic Ap stars in Figs.~\ref{fig:cold} and ~\ref{fig:hot}
are arranged roughly in order
of effective temperature.  The bottom spectra of
Fig~\ref{fig:hot} are
of the Am star, HR 1353 (above) and the F2 V $\sigma$
Boo (below).   Among these spectra, Figs. 3A and
especially 3B most closely resemble the upper schematic
profile of Fig. 1.  Figs. 2C, 2E, and 2F are examples
of the lower profile of Fig. 1.

\begin{figure*}
   \centering  
     \includegraphics[scale=0.9]{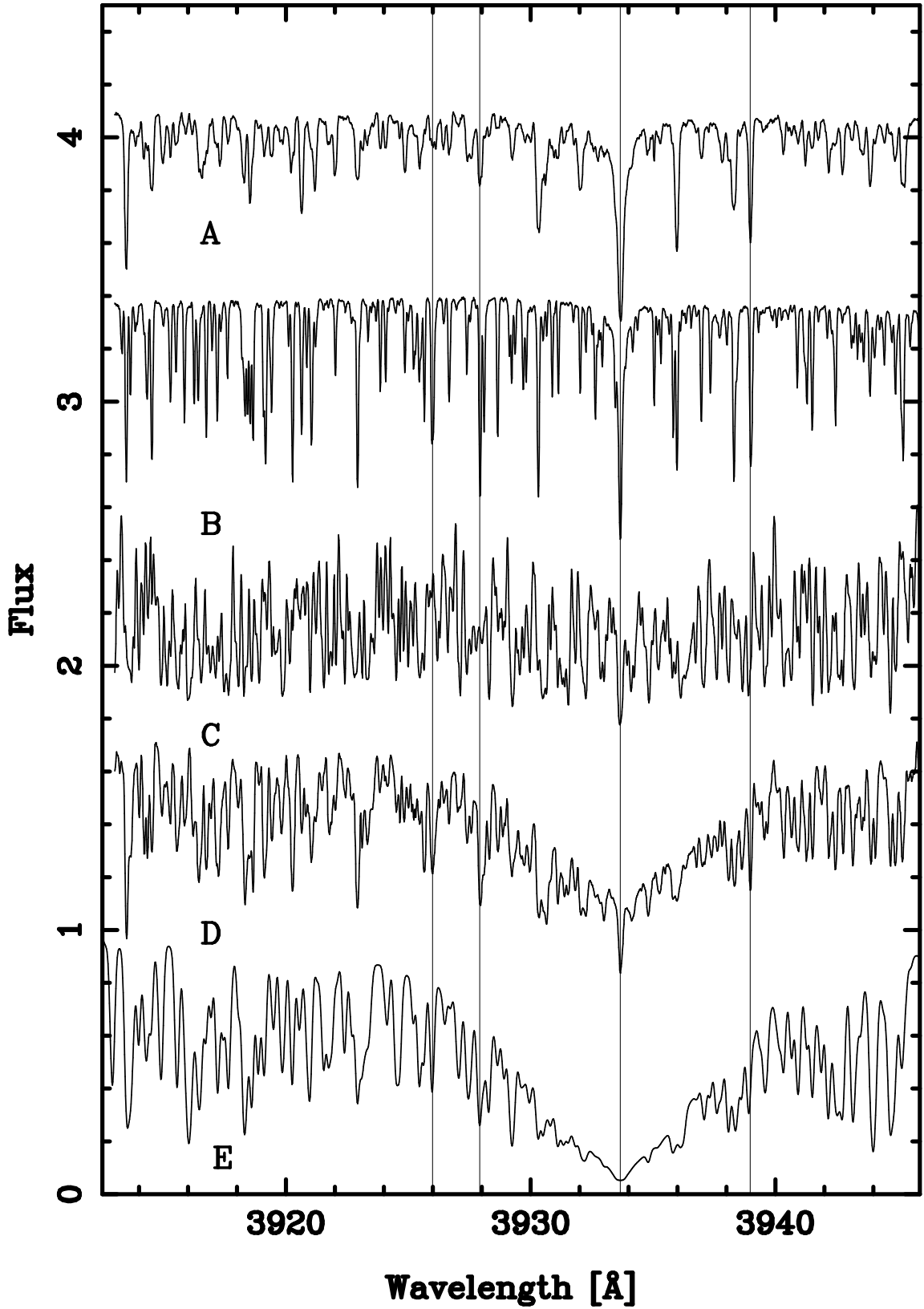}
      \caption{K-line region for A (HR 7552),
      B (HR 5623),
      C (Przybylski's star), D (HD 965), and
      E (a theoretical K-line profile).  Vertical lines as
      in Fig. 2.
         }
         \label{fig:misc}
   \end{figure*}

\placefigure{fig:misc}

Fig.~\ref{fig:misc} shows two hotter magnetic Ap stars (A and B),
 and the cool roAp,
Przybylski's star (C).  In all three cases, there is a deep,
sharp, core, but the wings are too weak to provide the
contrast we refer to as the wing-nib anomaly.  In the case
of Przybylski's star,
low-dispersion spectra
show broad, shallow absorption in the region of the
H+H$\epsilon$ and the K-line.  This absorption is not obvious in
Fig.~\ref{fig:misc}.  Nevertheless, one may say that
Przybylski's star
also shows a WNA, though the proportions of the near wing
and nib are
quantitatively different from those of the ten other cool
magnetic CP stars illustrated in Figs. 2 and 3.

HD 965 (Fig. 4D) has an intermediate temperature,
probably near 8000K.  \citet{bord03} 
discuss the difficulties
in finding a temperature for this star.
The lower plot (Fig. 4E) is an
LTE calculation using Michigan codes for a star with an
effective temperature\
of 6750K, $\log(g) = 4.0$, and enhanced abundances as in a
magnetic CP star.

Full nib widths, as defined in Fig.~\ref{fig:one}, are given
in Table~\ref{tab:nibwidths}.  The average in angstroms is
$0.42\pm 0.06$.  Provisional calculations show these nibs
are formed several thousand kilometers
above $\tau_{\rm 5000} = 1$.

\placetable{tab:nibwidths}
%
\section{Phase Variations}
\begin{figure}
   \centering  
  \includegraphics[width=0.3\textwidth,angle=270]{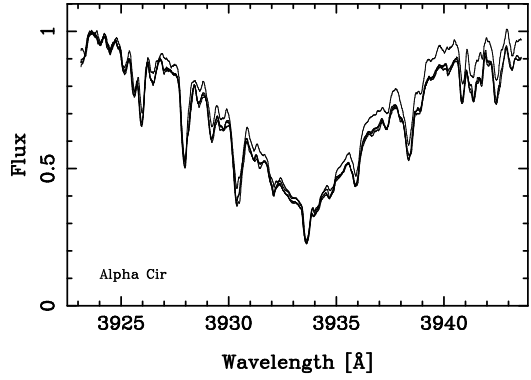}
      \caption{The K-line spectrum of $\alpha$ Cir at
      phases 0.19, 0.40, 0.50, and 0.94.  These are calculated
      from the rotational period of 4.463 days 
      \citep{kurtz94}.  Zero phase is JD 2449111.6852.
         }
         \label{fig:acir_op}
   \end{figure}

\begin{figure}
   \centering  
     \includegraphics[width=0.3\textwidth,angle=270]{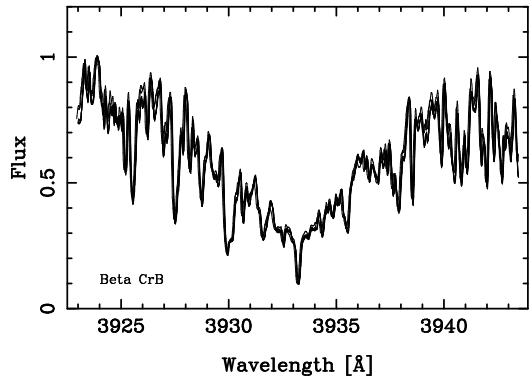}
      \caption{The K-line spectrum of $\beta$ CrB at
       phases 0.30, 0.60, 0.65, 0.76, and 0.92.
       These are calculated from the rotation period of
       18.4868 days \citep{kurtz89}.  
       Zero phase is JD 2434204.7.
         }
         \label{fig:bcrb_op}
   \end{figure}

The WNA would be washed out in the spectrum of a star with
significant rotational velocity.  Thus, the spectra
presented have all had sharp lines, and the stars themselves
relatively long rotational periods.  In Figs.~\ref{fig:acir_op}
and ~\ref{fig:bcrb_op}
we present overplots of the K-line region for two stars
with relatively shorter periods, $\alpha$ Cir and $\beta$ CrB.
Small variations are seen in both spectra, but the general
shape of the WNA persists.

\placefigure{fig:acir_op}

%

The relative constancy of these profiles might be compared
with those of Babel's (1992) Fig. 5 for 53 Cam, where
considerable phase variation is seen.  The 53 Cam spectra
appear to show WNA at phases 0.17 and 0.29.
These fundamental differences in phase behaviour show
the importance of
having bigger samples for the study of the WNA.
Any model that attempts to
explain this anomalous line profiles has to account for
variable as well as non-variable line profiles.

\placefigure{fig:bcrb_op}

\section{Non-CP stars}

\begin{figure*}[htp]
   \centering  
     \includegraphics[scale=0.9]{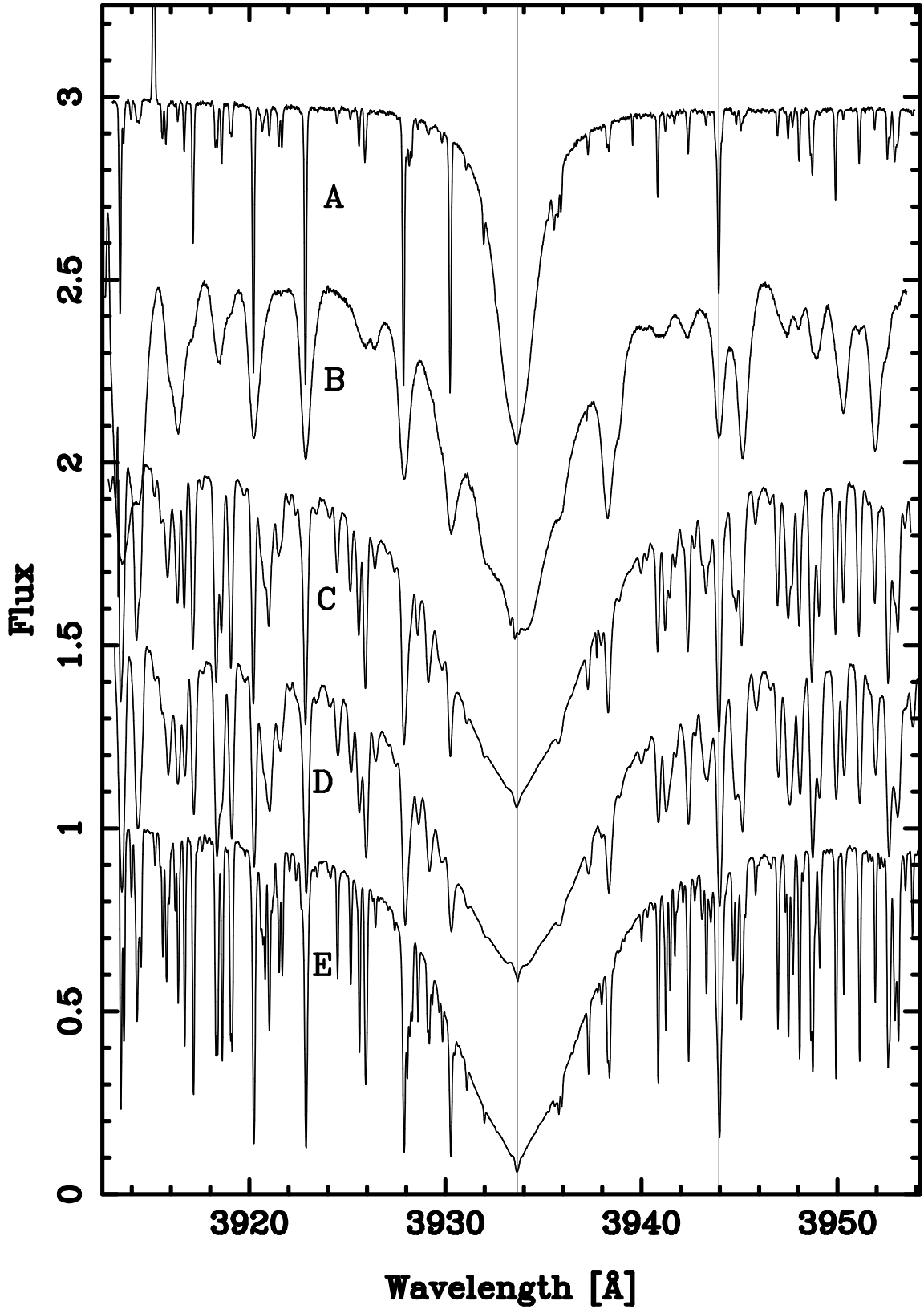}
      \caption{K-line region for A (HD 140283),
      B ($\gamma$ Dor),
      C (68 Eri), D (53 Vir), and
      E ($\gamma$ Pav).  Vertical lines as in
      previous figures.}
         \label{fig:non}
   \end{figure*}

\placefigure{fig:non}

We present K-line region spectra of several non-CP stars in
Fig.~\ref{fig:non}, to make clear the distinction of WNA profiles.
The upper spectrum is of the well-studied subdwarf, HD 140283.
The star has an effective temperature close to that of the sun,
but the lines are weak because of the very low metal-to-hydrogen
ratios.  Spectrum B is of $\gamma$ Dor, a pulsating
variable related to the $\delta$ Scuti stars, but surrounded by
a dust envelope 
\citep{balon94}.  The
bottom three spectra are of stars with MK types, though all
are metal poor; 68 Eri is F2 V, 53 Vir is F6 V, and $\gamma$ Pav
is F6 V.

\section{Discussion}
\begin{figure*}
   \centering  
      \includegraphics{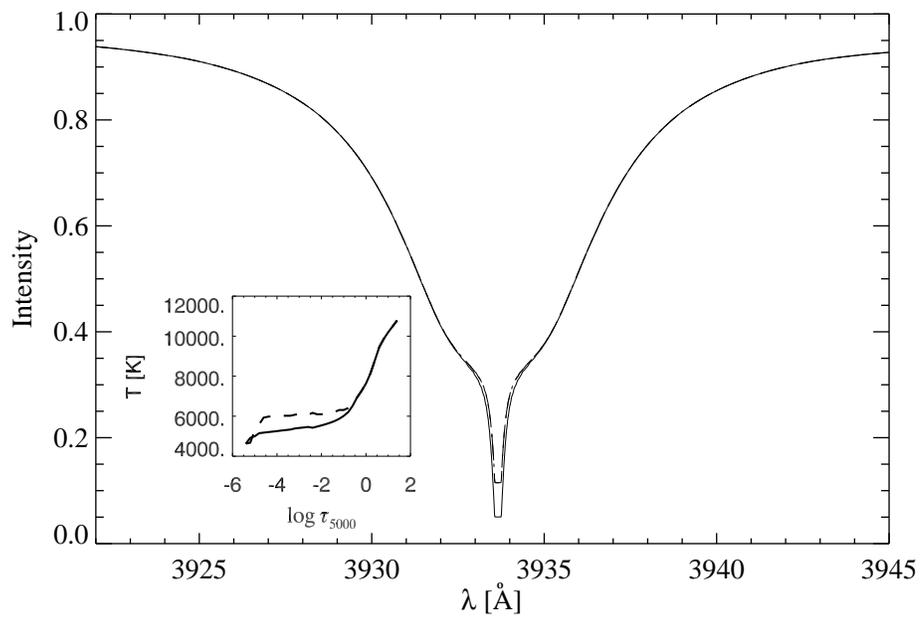}
      \caption{Ca\,{\sc ii}\,K in a cool star 
       (T$_{\rm eff}=6750$~K, $\log g=4.0$): LTE (solid line) and NLTE
       (dashed line) in the modified atmosphere. The inset shows the 
       original (solid line) and
       modified (dashed line) $T$--$\tau$ relation.
         } 
         \label{fig:bugger}
   \end{figure*}

\placefigure{fig:bugger}
It is our purpose to present an observational atlas
and not a new interpretation
of the wing-nib anomaly.  We do not dispute
stratification models.  They are demonstrably capable of
producing nibs.
Nevertheless, a few remarks are appropriate.

The
wing-nib and core-wing (Balmer lines) anomalies show
that the atmospheres of at least some of the cool CP stars
depart significantly from the classical models.  This was suggested
decades ago as an alternate explanation to abundance
anomalies---a way to explain the peculiar spectra with
normal abundances.
The objections were
dismissed forcefully by \citet{sarg66}.  
Briefly, he
argued that the CP stars had normal colors, curves of
growth, ionization temperatures, and Balmer profiles.
His arguments were valid at that time
though they are untenable today.  Nevertheless,
his conclusion is widely
accepted, that the photospheres of magnetic CP stars are
chemically anomalous.

Traditional abundance studies were based on classical,
plane-parallel atmospheres, with a uniform chemical
composition.
\citet{bab92} showed that a stratified
abundance structure would produce a Ca K-line nib while
Ryabchikova et al. (2002)
used a chemically stratified model similar to Babel's.
They also obtained a K-line nib, though they did not fit it
precisely.  They found certain other atomic lines that could
be fit more satisfactorily with the assumption of abundance
stratification than with a traditional model.

Much new work has been devoted to this phenomena.
Indeed, \citet{dwor05} 
described the
session of IAU Symposium 224 \citep{zver05}
on diffusion by saying ``the
keword ...was `stratification.'"
\citet{kochu02} showed that
the core-wing
anomaly of the Balmer lines might be explained by an ad hoc
variation in the $T$--$\tau$ relation.
We have made LTE calculations to show that wing-nib structure
of Ca {\sc ii} K may also be obtained with
a modified $T$--$\tau$.

Most of the calculations relevant for the WNA have
been made in LTE.  We have begun calculations
with the Kiel non-LTE code (Steenbock \& Holweger 1984)
that will be
reported in more detail 
elsewhere. We use the updated version of the Ca model
atom by \citet{wat85} \citep{stur93}.
The core of the Ca {\sc ii} K-line
forms in layers of the atmosphere, where both
involved levels show only very moderate deviations from
LTE ($<0.05$ dex).
Preliminary results show that these small quantitative
differences from LTE
in the K-line core vary somewhat with effective temperature.
As an example,
we present here a cool atmosphere (T$_{\rm eff}=6750$~K,
$\log g=4.0$)
with an ad-hoc modified $T$--$\tau$ relation (see inset of
Fig.~\ref{fig:bugger}). The atmosphere temperature was raised to
$T\sim 6000$~K over the optical depth range
$-4.5 < \log \tau < -0.5$.
Fig.~\ref{fig:bugger} shows the
resulting LTE (solid line) and NLTE (dashed line)
Ca\,{\sc ii}\,K profiles.
NLTE leads to a $\sim$ 5\% weaker core, but does not affect
the main profile
and the wings; thus, it would not lead to significant
abundance differences.
At present it seems that a fully non-LTE approach could
modify details of
a workable model, but would not require qualitative changes.

Neither of the above modifications takes account of the
influence of magnetic pressure on atmospheric structure.
Thus, it is reasonable to explore a third plausible departure
from a classical model atmosphere, based on a consideration
of magnetohydrodynamical effects.  We lack the information
necessary to take such effects explicitly into account.  One
possibility is that they simulate a model with gravity that
decreases outward \citep{conti65, valy04}.

Since a model with reduced gravity in the upper layers would
have lower pressured and therefore reduced atomic absorption,
the effect should resemble that of stratified
models.  We have made provisional calculations showing that
in LTE, such a model is indeed capable of yielding a K-line
nib.  Thus far, the modified-gravity models that give K-line
nibs have not yielded core-wing anomalous Balmer profiles.

\begin{acknowledgements}
CRC acknowledges useful conversations with H. A. Abt,
W. P. Bidelman, D. J. Bord,
and P. Hughes. We thank an anonymous referee for calling
our attention to the important studies of Babel.
Thanks are also due to the ESO staff for the UVESPOP
public data archive.  This research has made use of the
SIMBAD database, operated at CDS, Strasbourg, France.

{\it Facility:} \facility{VLT:Kueyen (UVES)}

\end{acknowledgements}

\clearpage

\begin{table}
\caption{Dates of Observations}
\label{tab:dates}
\begin{tabular}{rlcll}
\hline
\multicolumn{1}{c}{HD} &  
\multicolumn{1}{c}{Other} &  
\multicolumn{1}{c}{Date} &
\multicolumn{1}{c}{Phase}&
\multicolumn{1}{c}{Sp.}   \\   \hline
\object[HD 965]{965} &   BD +0 21    &   18/09/2002&    & A8p Sr   \\
\object[HD 24712]{24712} &   HR 1217     &   13/03/2001&    & A9 SrEuCr  \\
\object[HD 27290]{27290} & $\gamma$ Dor  &   19/09/2002&    & F4 III  \\
\object[HD 27411]{27411} &   HR 1353     &   14/08/2002&    & A3m  \\
\object[HD 33256]{33256} &   68 Eri      &   21/09/2002&    & F2 V  \\
\object[HD 101065]{101065} &   Przybylski's &   11/01/2003&    & Fp REE \\
\object[HD 114642]{114642} &   53 Vir      &   27/07/2001&    & F6 V     \\
\object[HD 116114]{116114} &BD -17 3829    &   06/02/2003&    & F0 SrCrEi \\
\object[HD 122970]{122970} &BD +6  2827    &   21/01/2002&    & F0p      \\
\object[HD 128167]{128167} &$\sigma$ Boo   &   13/03/2001&    & F2 V     \\
\object[HD 128898]{128898} &$\alpha$ Cir   &   25/01/2005&0.94& A3-A9    \\
       &               &   27/01/2005&0.19&     \\
       &               &   01/02/2005&0.40&     \\
       &               &   05/02/2005&0.50&     \\
\object[HD 133792]{133792} &   HR 5623     &   26/02/2002&    & A0 SrCr         \\
\object[HD 134214]{134214} & BD -13 4081   &   26/02/2002&    & F2 SrEuCr \\
\object[HD 137909]{137909} &$\beta$ CrB    &   03/02/2004&0.30& A9 SrEuCr    \\
       &               &   03/26/2004&0.60&     \\
       &               &   03/27/2004&0.65&     \\
       &               &   03/29/2004&0.76&     \\
       &               &   04/01/2004&0.92&     \\
\object[HD 137949]{137949} &   33 Lib      &   20/09/2002&F0p & SrEuCr \\
\object[HD 140283]{140283} & BD -10 4149   &   08/07/2001&    & sdF3     \\
\object[HD 176232]{176232} &   10 Aql      &   08/10/2001&    & A6 Sr    \\
\object[HD 187474]{187474} &   HR 7552     &   07/10/2001&    & A0 EuCrSi \\
\object[HD 188041]{188041} &   HR 7575     &   08/10/2001&    & A6 SrCrEu  \\
\object[HD 203608]{203608} &$\gamma$ Pav   &   06/12/2002&    & F6 V      \\
\object[HD 216018]{216018} &BD -12 6357    &   08/10/2001&    & A7 SrCrEu  \\
\object[HD 217522]{217522} &CD =45 14901   &   21/09/2002&    & A5 SrEuCr \\ \hline
\end{tabular}

\end{table}

\begin{table}
\caption{Measured Nib (Full) Widths}
\label{tab:nibwidths}
\begin{tabular}{rcrc}  \hline
star(HD) & width[\AA] & star(HD) & width[\AA] \\ \hline
965     &  0.40  & 137949 & 0.45 \\
24712   &  0.39  & 176232 & 0.33 \\
116114  &  0.40  & 188041 & 0.35 \\
122970  &  0.46  & 216018 & 0.39 \\
137909  &  0.45  & 217522 & 0.53 \\  \hline
\end{tabular}
\end{table}

%
%
\clearpage

%
%

%

\end{document}